\documentclass{JAC2003}

\usepackage{graphicx}
\usepackage{booktabs}

\begin{document}
\title{Beam heat load in superconducting wigglers}

\author{S. Casalbuoni\thanks{sara.casalbuoni@kit.edu},\\ ANKA, Karlsruhe Institute of Technology, Karlsruhe, Germany}

\maketitle

\begin{abstract}
   The beam heat load is a fundamental input parameter for the design of superconducting wigglers since
   it is needed to specify the cooling power. In this presentation I will review the possible beam heat
    load sources and the measurements of  beam heat load performed and planned onto the cold vacuum chambers
     installed at different synchrotron light sources.
\end{abstract}

\section{INTRODUCTION}

Superconducting (SC) wigglers are used worldwide in low and middle energy (1-3~GeV) storage rings to increase the flux in the harder part of the X-ray spectrum from 20 to 100~keV used for material science, biology, medical diagnostics and therapy~\cite{mezwallenSRN}.
In order to satisfy similar demands and further increase the brilliance SC undulators are under development for middle and high energy storage rings~\cite{yuriIPAC2012,scfls2011}. Free electron lasers would also benefit of superconducting undulators (elliptically polarised)~\cite{prestemonfel2011}. SC technology has been proposed also to be applied in undulators and wigglers for high energy physics projects as for the positron source of the International Linear Collider~\cite{clarke} and the damping wigglers for the Compact Linear Collider~\cite{schoerling}.

All these devices consist of a cryostat with SC NbTi coils kept at about 4~K and of a beam vacuum chamber to let the beam through the coils. The beam vacuum chamber also referred to as liner is kept at about 10-20~K. In order to maximize the peak magnetic field, the space between the liner and the coils should be minimized. Because of the necessity to impregnate the SC coils, they must be located out of the ultrahigh vacuum (UHV) where the beam is confined.  The liner intercepts the beam heat load and it is ideally thermally disconnected from the coils to avoid a degradation of their performance.  In reality the liner and the coils have always some thermal connection.
A proper cryogenic design of all the devices described above requires the knowledge of the beam heat load to the beam vacuum chamber.
In the following section I describe some of the possible beam heat load sources. I then report on the measurements of the beam heat load to cold vacuum chambers performed at different synchrotron light sources with the installed SC undulators and wigglers. Afterwards I present dedicated experiments to measure the beam heat load to a cold vacuum chamber which will hopefully be useful also to understand the underlying mechanism. The last section contains conclusions and outlook.

\section{POSSIBLE BEAM HEAT LOAD SOURCES}

Possible beam heat load sources are: synchrotron radiation, RF effects
due to geometrical and resistive wall impedance, and electron and/or ion bombardment.

\subsection{Synchrotron Radiation Heating}
\begin{figure}[h]
\centering
\includegraphics[width=86mm]{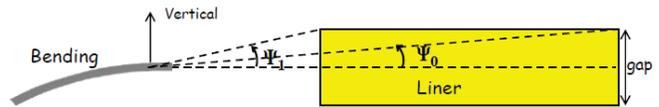}
  \caption{Scheme of the synchrotron radiation from the upstream bending magnet hitting  the upper
and lower surfaces of the liner of a SC undulator or wiggler.}
  \label{synchrotron}
  \end{figure}

The power of the synchrotron radiation emitted from the upstream bending magnet hitting the upper and lower surfaces of the vertical gap of the SC undulator or wiggler is~\cite{wallen,scPRSTAB2007}:

\begin{equation}
P_{\rm syn}=2P_0\frac{21}{32}\int_{\psi_0}^{\psi_1}
\frac{\gamma}{(1+\gamma^2\psi^2)^{5/2}}\Biggl[1+\frac{5}{7}\frac{\gamma^2\psi^2}{(1+\gamma^2\psi^2)}\Biggr] d\psi
\end{equation}
where $\psi_0$ and $\psi_1$ are the lower and upper values of $\psi$ indicated in Fig.~\ref{synchrotron},
$$\gamma=E/m_ec^2~,$$
$$P_0= \frac{eI\gamma^4}{ 6\pi \epsilon_0 \rho},$$
$e$ is the electron charge, $I$ is the average beam current, $\epsilon_0$ is the vacuum permittivity, $E$ is the beam energy, $\rho$ is the radius of curvature of the electron trajectory in the bending magnet, $m_e$ is the electron mass and $c$ is the speed of light.
The factor two in front of the integral takes into account of the upper and lower surfaces of the vacuum chamber of the SC undulator or wiggler (see Fig.~\ref{synchrotron}).

The beam heat load contribution from synchrotron radiation depends linearly on the stored average beam current, it depends on the electron beam energy and on the geometry, that is on the relative position of the bending magnet, of the collimator and the liner. It is however independent on the filling pattern and on the bunch length.

\subsection{RF Heating}

\begin{figure}[h!]
\center
\includegraphics[width=85mm]{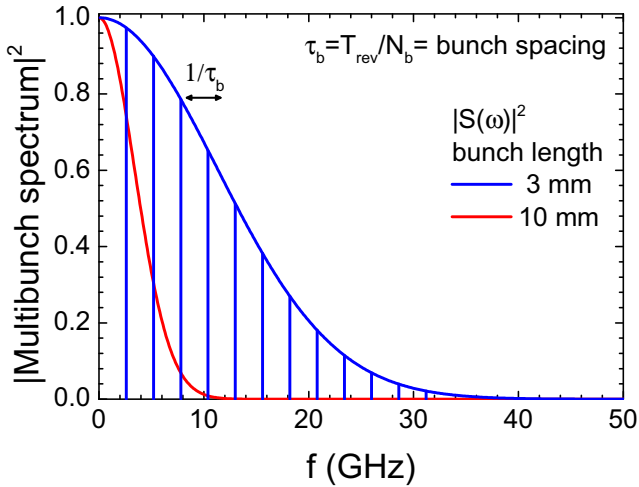}
  \caption{Single bunch spectrum with two different bunch lengths (solid red and blue lines) and sketch of multibunch spectrum (solid blue vertical lines modulated by the single bunch spectrum)~\cite{JINST}.}
 \label{beamspectrum}
\end{figure}

The total power $P_{RF}$ lost by the beam due to the wake fields of $N_b$ equally spaced bunches can be obtained by using the relation~\cite{heifets99}:
\begin{equation}
P_{RF}= I^2\sum\limits_{n=-\infty}^{\infty}ReZ_{||}(nN_b\omega_0) \left|S(nN_b\omega_0)\right|^2
\label{PowerRFsum}
\end{equation}
$S$ being the single bunch spectrum, $ReZ_{||}$ the real part of the longitudinal component of the coupling impedance.
Assuming bunches with Gaussian shape and length $\sigma_z=3$~mm a schematic representation of the multibunch spectrum not in scale is shown by the blue vertical lines in Fig.~\ref{beamspectrum}. If we would plot the lines spaced in scale they would be indistinguishable from the single bunch spectrum. As obtained from Eq.~(\ref{PowerRFsum}), when $N_b$ times the revolution frequency $f_0$ (for a 300~m circumference storage ring $f_0=\omega_0/(2\pi)\sim 1$~MHz) is much smaller than the inverse of the bunch duration $c/(\sqrt2 \pi\sigma_z)\sim$~few 10 GHz ($c=$ speed of light), the multibunch spectrum is well approximated by the single bunch spectrum $$S(\omega)=\exp{-\frac{\sigma^2_z\omega^2}{2c^2}}.$$
In this case and in absence of resonant modes $N_b\omega_0\rightarrow d\omega$ and $nN_b\omega_0\rightarrow \omega$, so Eq.~(\ref{PowerRFsum}) becomes:
\begin{equation}
P_{RF}=\frac{I^2 T_0}{N_b}k_l
\label{PowerRFint}
\end{equation}
where $k_l$ is the loss factor, $T_0=2\pi/\omega_0$ the revolution period and $I=N_bQ/T_0$ the average beam current with $Q$ the average bunch charge.
The loss factor is given by:
\begin{equation}
k_{l}=\frac{1}{\pi}\int\limits_{0}^{\infty} \left|S(\omega)\right|^2ReZ_{||}(\omega)d\omega~.
\label{KL}
\end{equation}
The total power $P_{RF}$ lost by the beam is an upper limit of the power dissipated in the structure since, excluding the case of resistive wall, it is unknown where this power is deposited.
In case of resistive wall the power is deposited in the first few $\mu$m of the vacuum chamber. For wakes induced by geometrical changes of the cross section of the vacuum chamber the power could be deposited in the chamber itself, or could be exchanged in the interaction with other bunches and be deposited somewhere else in the accelerator~\cite{JINST}.

Geometrical changes in the cross section are almost unavoidable at the transitions, where tapers and RF fingers and bellows are employed. Even flanges connecting parts with the same aperture contribute to cross section changes due to the finite mechanical accuracy of the manufactured components. Since however these parts are far away from the coils, towards the entrance and exit of the cryostat and thermally anchored to a radiation shield at $\sim 50-80$~K, their contribution to heat the central liner close to the coils is expected to be negligible.
In case of high ordered modes excited by the beam and trapped in the liner it should be possible to considerably remove the losses by changing the inverse of the bunch spacing by the bandwidth of the resonance~\cite{spataro}.

The contribution of the surface roughness to the impedance is relevant for bunches with a length of the order of magnitude of the surface corrugations.
The different theoretical models developed to calculate the coupling impedance of a beam pipe with a rough surface are reviewed in Ref.~\cite{revmostacci,revbane}.  An important parameter to determine the impedance is the so called aspect ratio, that is the ratio between the average peak heights and the average distance between the peaks.

In order to reduce the losses due to resistive wall heating the material chosen for the surface ( of few $\mu$m) of the liner exposed to the beam is a high conductivity material as copper. Aluminum is also used, for example in the SC undulator under development at the Argonne Photon Source (APS)~\cite{yuriIPAC2012}.
The real part of the longitudinal impedance due to resistive wall is given by:
\begin{equation}
ReZ_{||}(\omega)=\frac{L}{\pi 2b}R_{surf}
\label{RWH}
\end{equation}
where $L$ is the length of the considered portion of vacuum chamber, $R_{surf}(\omega)$ is the surface resistance, and in case of a circular beam pipe $2b$ is the diameter~\cite{chou} while in case of a rectangular beam pipe is the gap~\cite{wallen}.
 For copper at low temperatures and $RRR>7$ the anomalous skin effect~\cite{wallen, scPRSTAB2007, chou, london} has to be considered:
\begin{equation}
R_{surf}(\omega)=R_{\infty}(\omega)(1+1.157\alpha^{-0.276}),~~~ \mbox{for}~\alpha \ge 3
\end{equation}
with
$$\alpha=\frac{3}{2}\left[\frac{\ell}{\delta(\omega)}\right]^2=\frac{3}{4}\mu_r \mu_0 \sigma \omega \ell^2$$
where $\ell$ is the mean free path,
$$\delta(\omega)=\sqrt{\frac{2}{\mu_r\mu_0\sigma \omega}}$$
 the skin depth, $\mu_r$ the relative permeability, $\mu_0$ the vacuum permeability and $\sigma$ the electrical conductivity at room temperature (for copper $\sigma=6.45\times 10^7$~S/m), and with
$$R_{\infty}(\omega)=\left(\frac{\sqrt{3}}{16\pi}\frac{\ell}{\sigma}(\mu_r\mu_0\omega)^2\right)^{1/3}~.$$

The beam heat load due to RF effects depends quadratically on the stored average beam current. It depends on the bunch length and on the filling pattern, in particular on the number of bunches and on the bunch spacing, and on the position of the bunch in the vacuum chamber. It does not depend on the beam energy.

\subsection{Electrons and/or Ions Bombardment Heating}

Ions and electrons created by ionization and photodesorption, and accelerated against the wall by the passing beam will also contribute to the beam heat load. The beam dynamics involved is unknown and might be quite complicated. It is however likely that it is dominated  by the beam properties and by the chamber surface characteristics, as secondary emission yield, photoemission yield, photoemission induced electron energy distribution, etc..., which are only partially measured for a cryosorbed gas layer.

Since the beam dynamics is unknown we do not know for this source of beam heat load its dependence on the different beam parameters as filling pattern, beam energy, average stored beam current, bunch length, bunch spacing and number of bunches. Taking this into account we cannot then state that an observed linear or quadratic dependence of the beam heat load on the average beam current is sufficient to prove that the main contribution to the beam heat load comes from synchrotron radiation or RF effects, respectively.

\section{OBSERVATIONS WITH SC WIGGLERS AND UNDULATORS}

Cold bore SC wigglers and an undulator installed in different storage rings have been used
also to measure the beam heat load. The interpretation of the measurements is not straightforward since
these devices have not been designed to perform beam heat load diagnostics. In all cases the beam heat load measured is higher than the one expected from the synchrotron radiation of the upper bending magnet and from resistive wall heating.
In the following I summarize the results from the measurements performed with SC wigglers at MAX II, at the Diamond Light Source (DLS), and with a SC undulator at ANKA (\AA ngstrom source Karlsruhe).

\subsection{Experience at MAX II: SC Wiggler}
The two SC wigglers designed and manufactured at MAX-lab successfully operating for almost a decade showed both a higher helium consumption than predicted~\cite{mezwallenSRN}.
For one of the wigglers the beam heat load has been measured to be 0.86~W instead of the predicted 0.17~W. The beam heat load measured as a function of the stored average beam current, can be fitted by the sum of a linear and of a quadratic component, respectively made responsible of synchrotron radiation and resistive wall losses.
The contribution from the synchrotron radiation is double than the one predicted. This discrepancy has been attributed to a misalignment of the bending magnet-collimator-liner system.
The contribution to the beam induced heating from the image currents is 0.59~W, about 10 times larger than expected from the calculations~\cite{wallen}, is not understood.

\subsection{Experience at DLS: SC Wigglers}
Two SC wigglers from the Budker Institute for Nuclear Physics are installed at the DLS. The beam heat load is extrapolated  by using
the temperature rise in the liner and the heat shields to deduce the extra cooling power of the cryocoolers plus
the additional liquid helium boil off~\cite{schoutenIPAC11}. The uncertainty in the measurements is up to $30\%$.
A quadratic dependence on the bunch charge and on the stored average beam current is observed, and also in this case the predicted values are smaller than the measured ones.

\subsection{Experience at ANKA: SC Undulator}
A cold bore superconducting undulator built by ACCEL Instr. GmbH, Bergisch Gladbach, Germany~\cite{scPRSTAB2006}, was
installed in one of the four straight sections of the ANKA storage ring in March 2005 and removed in July 2012. The performance of this device was limited by the too high beam heat load. Namely, the superconducting coils performance was reduced during users operation from 750~A to 300~A meaning a reduction in the peak magnetic field on axis from 0.42~T to 0.26~T. The observed beam heat load up to 2.5~W~\cite{scIPAC12} at a gap of 8~mm and at 100~mA stored average beam current is much higher than the predicted values of 63~mW from the synchrotron radiation of the upstream bending magnet and of 22~mW from the image currents~\cite{scPRSTAB2007}.

A simple model of electron bombardment appears to be consistent with the large variation of beam heat load and of pressure rise values as a function of the average beam current for different gaps~\cite{scPRSTAB2007,scIPAC12} observed in the cold bore of the SC undulator. Still to be understood is the mechanism responsible for the electron multipacting and the role played by the cryosorbed gas layer.
A common cause of electron bombardment is the buildup of an electron cloud, which strongly depends on the chamber surface properties. The
surface properties as secondary electron yield, photoemission yield, photoemission induced electron energy distribution, needed in the simulation codes to determine the possible occurrence and size of an electron cloud buildup, have only partly been measured for a cryosorbed gas layer. Even using uncommonly large values for these parameters, the heat load inferred from the ECLOUD simulations~\cite{ecloud} is about one order of magnitude lower than the measurements~\cite{iriso}.
While electron cloud buildup models have been well benchmarked in machines with positively
charged beams, in electron machines they do not reproduce the observations satisfactory. This has been shown at the ECLOUD10 workshop also by K. Harkay~\cite{harkay} and by J. Calvey~\cite{calvi} comparing
the RFA data taken with electron beams in the APS and in CesrTA, respectively, with the simulations performed using the electron cloud buildup codes POSINST~\cite{posinst} and ECLOUD~\cite{ecloud}. From these comparisons it seems that the electron cloud buildup codes do not contain all the physics going on for electron beams. In order to fit the data with the simulations, the approach at APS and CesrTA is to change the photoelectron model. At ANKA we tried to study if the presence of a smooth ion background (i.e. a partially neutralized electron beam) can change the photoelectron dynamics so that the photo-electrons can receive a significant amount of kinetic energy from the ion cloud plus electron beam system. Following preliminary analytical results by P. F. Tavares (MAX-lab), S. Gerstl (ANKA) has included an ion cloud potential in the ECLOUD code: preliminary simulations are encouraging.

\section{Dedicated experiments}

\subsection{LBNL-SINAP Calorimeter}
\begin{figure}[h!]
\center
\includegraphics[width=80mm]{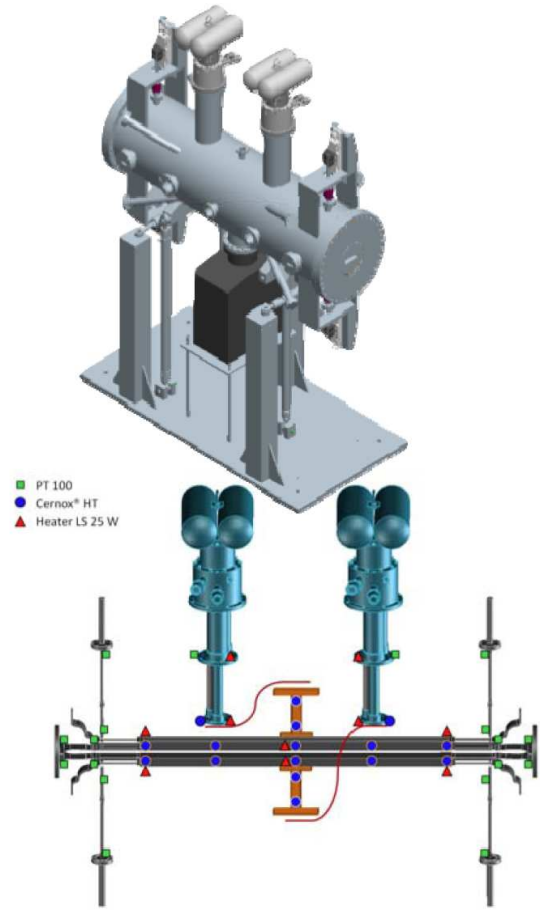}
  \caption{Sketch of the LBNL-SINAP calorimeter~\cite{prestemonfel2011}.}
 \label{calorimeter}
\end{figure}

A calorimeter to measure the beam heat load in a storage ring  via temperature gradients has been proposed by the Lawrence Berkeley National Laboratory (LBNL)~\cite{Trillaud}. Two proposals with different cooling concepts have been made: one using a He boiler and the other conduction cooling. This last concept will be realized in collaboration with the Shanghai Institute of Applied Physics (SINAP) and the device is planned to be installed in the Shanghai Light Source~\cite{prestemonfel2011}.
The LBNL-SINAP calorimeter, shown in Fig.~\ref{calorimeter}, will allow to measure the beam heat load at different gaps. It will be provided with heaters to permit constant temperature operation and in situ calibration checks. Measurements for different materials will be possible by changing the substrate of the liner which faces the beam.

\subsection{COLDDIAG}

With the aim of measuring the beam heat load on a cold bore and in order to gain a deeper understanding in the beam heat load mechanisms, a cold vacuum chamber for diagnostics (COLDDIAG) has been proposed~\cite{scepac08} and built~\cite{scasc10}. This project led by ANKA is in collaboration with CERN, DLS, Frascati National Laboratory, Rome University ``La Sapienza", STFC Daresbury Laboratory, STFC Rutherford Appleton Laboratory, University of Manchester, Cockcroft Institute of Science and Technology and Lund University MAX-lab. The vacuum chamber is being designed and fabricated in collaboration with Babcock Noell GmbH.

\begin{figure}[h!]
\center
\includegraphics[width=85mm]{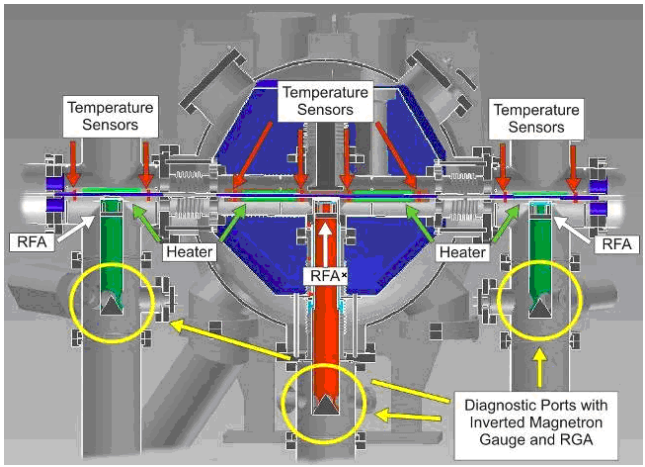}
  \caption{Overview of the cryostat and the diagnostics
installed in COLDDIAG.~\cite{sgipac12}.}
 \label{colddiag}
\end{figure}

COLDDIAG consists of a cold vacuum chamber (see cryostat in Fig.~\ref{colddiag}) located between two warm sections. This will allow to observe the influence of synchrotron radiation on the beam heat load and a direct comparison between the cryogenic and room temperature regions, with and without a cryosorbed gas layer, respectively. The same suite of diagnostics is used in both the cold and warm regions. The diagnostics being implemented are: i) retarding field analyzers to measure the electron flux, ii) temperature sensors to measure the total heat load, iii) pressure gauges, iv) and mass spectrometers to measure the gas content. In addition, to suppress charged particles from hitting the chamber wall a solenoid is installed on the downstream half of the cold liner section. The magnet reaches on axis a magnetic field of around 10~mT with a current of 1~A. The inner vacuum chamber will be removable in order to test different geometries and materials. COLDDIAG is built to fit in a short straight section at ANKA, but ANKA is proposing its installation in different synchrotron light sources with different energies and beam characteristics.

A successful final acceptance test has been performed with the liner reaching a temperature of 4 K and the beam vacuum a pressure of 10$^{-9}$ mbar~\cite{sgipac11}.

COLDDIAG was installed in the storage ring at the DLS in November 2011. Due to a mechanical failure of the thermal transition of the cold beam tube, the cryostat had to be removed after one week of operation. Preliminary results show a quadratic behaviour of the beam heat load as a function of average beam current. The measured beam heat load of $\sim$8~W at 250~mA is almost two orders of magnitude larger than the predicted value from resistive wall heating $\sim0.1-0.2$~W. Even if more statistics is needed, the almost random temperature distribution on the liner and the small but visible effect of the solenoid on the temperature distribution point out to electron bombardment as at least one component of the beam heat load observed~\cite{sgipac12}. Currently the design of the liner thermal transition is changed and a second installation at the Diamond Light Source is under discussion.

During a longer installation in the DLS it is planned to monitor the temperature, the electron flux, the pressure and the gas composition changing~\cite{scasc10}:

\begin{itemize}
\item the average beam current to compare the beam heat load data with synchrotron radiation and resistive wall heating predictions,
\item the bunch length to compare with resistive wall heating predictions,
\item the filling pattern in particular the bunch spacing to test the relevance of the electron cloud as heating mechanisms,
\item beam position to test the relevance of synchrotron radiation and the gap dependence of the beam heat load,
\item inject different gases naturally present in the beam vacuum (H$_2$, CO, CO$_2$, CH$_4$) to understand the influence of the cryosorbed gas layer on the beam heat load, and eventually
identify the gases to be reduced in the beam vacuum.
\end{itemize}

\section{CONCLUSIONS AND OUTLOOK}

The beam heat load measurements performed with cold bore SC wigglers and an undulator installed in different storage rings are not yet understood.

Two upcoming dedicated experimental setups, the LBNL-SINAP calorimeter and the COLDDIAG, will be able to measure the beam heat load with high accuracies $< 0.05$~W and
hopefully help to understand the beam heating mechanism. Even if both setups are designed to measure the beam heat load, they are nicely complementary. While the LBNL-SINAP calorimeter
will allow beam heat load measurements at different gaps, the COLDDIAG has one cold and two warm sections, and it is equipped with additional diagnostics as retarding field analyzers, pressure gauges and mass spectrometers to shed light on the role played by the cryogenic layer in the beam heating mechanism. Preliminary measurements performed with the COLDDIAG installed at the DLS indicate a value of the beam heat load of $\sim 8$~W at 250~mA, which is almost two orders of magnitude larger than the predicted value from resistive wall heating $\sim$ 0.1 - 0.2~W.

Additional studies on the beam heat load can come from the SC wigglers installed in many different storage rings and from the new SC undulators to be installed at ANKA and at the APS.

\section{ACKNOWLEDGMENT}
I would like to thank M. Migliorati, A. Mostacci (University of Rome ``La Sapienza" and LNF, Frascati, Italy) and B. Spataro (LNF, Frascati, Italy) for useful discussions on RF heating.

\end{document}